\documentclass[aps,prb,twocolumn,superscriptaddress]{revtex4-1}

\usepackage{graphicx}
\usepackage{textcomp}
\usepackage{lmodern}

\begin{document}

\title{Geometry-induced reduction of the critical current in superconducting nanowires}

\author{D. Henrich}
\email[]{dagmar.rall@kit.edu}

\author{P. Reichensperger}

\author{M. Hofherr}

\author{J. M. Meckbach}

\author{K. Il'in}

\author{M. Siegel}
\affiliation{Institut f\"ur Mikro- und Nanoelektronische Systeme (IMS), Karlsruher Institut f\"ur Technologie, Hertzstrasse 16, 76187 Karlsruhe, Germany}

\author{A. Semenov}
\affiliation{DLR Institut f\"ur Planetenforschung, 12489 Berlin, Germany}

\author{A. Zotova}

\author{D. Yu. Vodolazov}
\affiliation{Institute for Physics of Microstructures, Russian Academy of Sciences, 603950, Nizhny Novgorod, GSP-105, Russia}

\date{\today}

\begin{abstract}
Reduction of the critical current in narrow superconducting NbN lines with sharp and rounded bends with respect to the critical current in straight lines was studied at different temperatures. We compare our experimental results with the reduction expected in the framework of the London model and the Ginsburg-Landau model. We have experimentally found that the reduction is significantly less than either model predicts. We also show that in our NbN lines the bends mostly contribute to the reduction of the critical current at temperatures well below the superconducting transition temperature.
\end{abstract}


\maketitle

\section{Introduction}

The experimental realization of the depairing critical current in small superconducting specimens was a subject of intense efforts since the beginning of the 60's. The problem was solved by showing that in a one dimensional wire or strip the depairing critical current can be reached when certain experimental care is taken \cite{1Bar62,1Sko76,1Kup80,1Ger80,1Neu82,1Rom82}. Specifically, in a thin strip with a width less than the magnetic penetration depth, the depairing current may be reached if magnetic vortices cannot enter the strip. Although a potential barrier exists that hampers permanent vortex entry \cite{1Bean64}, there is a non-zero thermodynamic probability that a vortex will be thermally excited over the barrier and appears in the strip \cite{1Taf06}. In this case the Lorentz force caused by the current will steer the vortex to either side of the strip. Strong enough pinning may freeze vortices in the strip until the current becomes sufficiently large in order to tear them off from the pinning centers. In case of sufficiently strong pinning the critical state can still be caused by depairing. However the measured value of the critical current will be lower than the theory predicts due to a decrease of the superconducting cross-section by normal cores of the pinned vortices. In narrow strips with widths less than $4.4 \xi$, where $\xi$ is the superconducting coherence length, vortices cannot exist \cite{1Lik97}. In this case the depairing critical current can be realized regardless of the pinning centers in the specimen. The magnitude of the depairing current depends on the purity of the superconductor and the degree of localization. For extremely pure or dirty superconductors in the local limit analytical expressions for the critical current near the superconducting transition can be derived, while for intermediate cases as well as for the whole temperature range only numerical evaluations provide the correct values \cite{1Kup80}. 

For a variety of superconducting electronic applications, the highest achievable current is an important aspect. For example, the detection efficiency of a superconducting nanowire single photon detector (SNSPD) \cite{1Gol01} grows when the operation current increases \cite{1Haas07}. Therefore, knowledge of the nature of the experimentally measured critical current is essential for device optimisation. Although the two-fluid type of the temperature dependence of the critical current was often observed in earlier experiments \cite{1Ger80,1Neu82}, this fact alone does not justify that the depairing current has been realized. Even if the line width remains unchanged, any deviation from the straight-line geometry further reduces the measurable critical current due to current crowding and corresponding local increase in the current density near bends or curves. In order to fill an area larger than the optical wavelength, the nanowire in SNSPD is usually patterned as a meander with sharp 180$^{\circ}$ turns. The thickness $d$ and the width $w$ of typical SNSPD nanowires satisfy the conditions $d\leq \xi$ and $w \ll \lambda_{\text{eff}}$ where $\lambda_{\text{eff}} = \lambda^2 d^{-1}$ and $\lambda$ is the magnetic penetration depth. Therefore the current density remains uniform over the wire cross-section in the straight portions of the meandering nanowire. The uniformity is disrupted at the turns where the current crowds at the inner edges. The crowding increases the local current density above the mean current density in the straight portions of the wire. Therefore, the potential barrier for the vortex entry first disappears near the turns and this decreases the measurable critical current in comparison to the straight wire of the same cross-section. Another consequence of the current crowding near turns is that the practically achievable ratio of the operation current to the depairing current in the straight portions remains less than it could be in a wire without turns. This makes straight portions less effective in detecting photons at wavelengths where vortex-assisted detection \cite{2Hof10} takes place. Hence, decreasing the strength of current crowding near turns would greatly improve the spectral range of these detectors.

The influence of turns with various shapes on the critical current has been studied theoretically in the framework of the London equations for the transport current and the screening current around a vortex \cite{1Cle11}. Another approach, which we also involve for comparison here, is the numerical solution of the Ginsburg-Landau (GL) equation \cite{1Tin96} for the superconducting line with turns. This allows one to find the current at which the superconducting state becomes unstable. We compare predictions of these two approaches with the experimentally observed reduction of the critical current in superconducting lines with different bends. We show that in qualitative agreement with both theories the reduction increases with the increase in the bending angle and the decrease in the arc radius of the inner edge of the bend. However, the amount of the reduction and its temperature dependence noticeably differ from theoretical predictions. 

\section{Experiment}

We have studied the effect of bends on the critical current of superconducting NbN lines which had nominal width of 300\,nm and a thickness close to 10\,nm. In SNSPDs from NbN films, the widths and thicknesses of nanowires are typically less than 100 and 5\,nm, respectively. The enhanced cross-section of our lines greatly improved reproducibility of fabrication while it still satisfied typical SNSPD conditions $d \leq \xi$ and $w \ll \lambda_{\text{eff}}$. Also, the higher critical currents significantly simplified measurements and improved experimental accuracy. Moreover, since the amount of the reduction in the critical current due to bends is lager for structures with larger $w/\xi$ ratio, an increased width allows for easier observation and analysis of this effect. The samples had layouts as shown in insets in Figs.~\ref{fig1} and \ref{fig2}. They included two symmetric bends connected by portions of straight lines to each other and to enlarged contact pads. Tapered connections to contact pads decreased current crowding near the contacts. Samples from the batch \#1 consisted of lines with two sharp bends whose bending angle $\theta$ varied from $\theta = 90^{\circ}$ to $\theta = 0^{\circ}$; the latter angle corresponds to a straight line. Batch \#2 consisted of lines with two rounded $90^{\circ}$ bends. Rounding was made by arc segments with different inner radii $r$, which varied from just a few to 500\,nm, and the outer radii $r + w$, correspondingly. In order to reduce the dispersion of the experimental data, we prepared and tested several (typically 3 to 4) nominally identical samples with each layout. To analyse the effect of bending angle and radius at 4.2\,K we used the critical current densities averaged over all samples with the identical layout.

The resistive superconducting transition was measured for each sample at a bias current less than 1\,\textmu A. We defined the transition temperature $T_C$ as the temperature corresponding to the 0.1\% of the normal state resistance at 20\,K. The critical currents were measured in the temperature range from $T_C$ to 4.2\,K in the current-bias mode. We used slow, a few second long current sweeps and battery driven electronics with reduced noise level. This regime mimics the operation of SNSPDs and is relevant if results are to be used for further detector development. However, it is subject to electromagnetic interference and fast fluctuations whose effect reduces the accuracy of determining the critical current near the transition temperature. At temperatures $T > 0.95 T_C$, where the current-voltage characteristics were non-hysteretic, the critical current $I_C$ was determined using a 100\,\textmu V criterium emerging from the voltage resolution of our experimental setup. At lower temperatures $I_C$ was defined as the bias current at which bridges showed clear jump from the superconducting to the normal state.

\subsection{Technology}
The films were deposited on R-plane optically polished sapphire substrates by reactive magnetron sputtering of a pure Nb target in a gas mixture of Argon and Nitrogen. The substrates were kept at 750$^{\circ}$C during film deposition. Details on the NbN thin film deposition and characterization of the films can be found in ref. \onlinecite{1Hen12*}. The lines were defined by electron beam lithography and consecutive reactive ion etching in a parallel plate reactor.
To find the most reliable parameters for the patterning process, they were optimized on 9\,nm thick NbN samples structured into 130\,nm wide stripes. As etching gases, pure SF$_6$ as well as SF$_6$ in combination with Ar or O$_2$ with varying flow ratios, gas pressures and etching power of the reactive ion etching process in conjunction with the electron beam exposure parameters have been investigated. The criteria for the optimization were a minimal reduction of $T_C$ in respect to the unstructured film and a minimal spread in the resistance around the nominal value. For processes fulfilling these criteria, they were further optimized for the maximal $j_C$-values with the highest reproducibility. The optimal conditions were achieved for a SF$_6$:Ar mixture with 3:1 flow ratio, a total pressure of 90\,mTorr, an etching power of 100\,W and 90\,s etching time.
Those parameters were then used to structure the 300\,nm wide nanowire samples with the bends. In a consecutive step, the contact pads were made from the same film by means of photolithography and the same etching technique. The mean line thickness $d = 10.4$\,nm was obtained by a step-profilometer as an average of several measurements at different locations along the line. The actual line-width in our samples varied between 250 and 300\,nm; it was measured for each sample with an accuracy better than 10\,nm with a scanning electron microscope.

\subsection{Results}
The superconducting transition temperatures of all studied samples were spread over the interval 13.1 - 13.6 K. However we did not observe any correlation of the $T_C$ values with either the bending angle or the radius of the connecting arcs.

\begin{figure}
\includegraphics{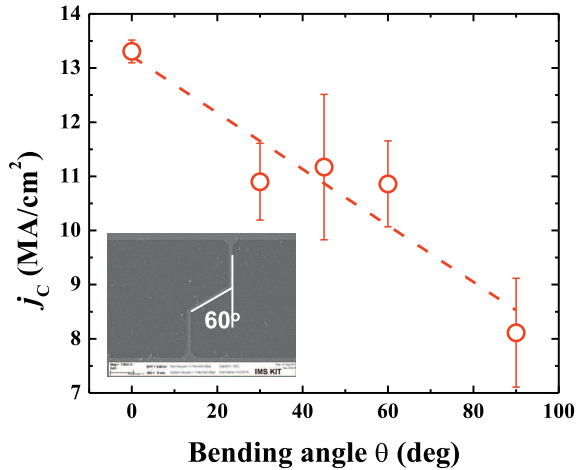}
\caption{\label{fig1}(Color online) Averaged critical current densities at 4.2\,K for samples with different bending angles. Zero angle corresponds to the straight line. Inset shows the sample layout for the 60$^{\circ}$ angle.}
\end{figure}

\begin{figure}
\includegraphics{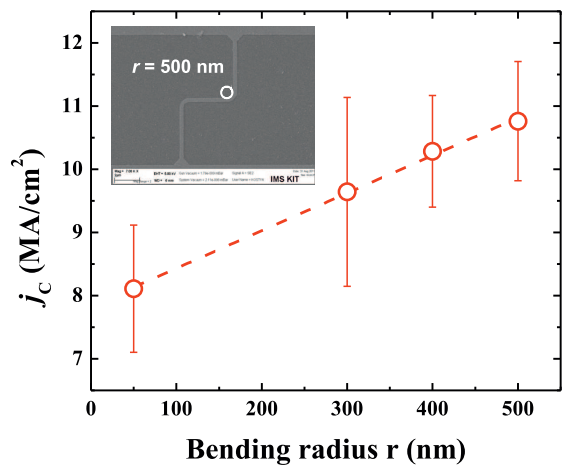}
\caption{\label{fig2}(Color online) Averaged critical current densities at 4.2\,K for samples with two 90$^{\circ}$ bends and different inner radii of the connecting arc. The inset shows the sample layout with the arc radius 500\,nm.}
\end{figure}

Figure \ref{fig1} illustrates the effect of the bending angle on the measured critical current for samples from the batch \#1. The critical current density $j_C$(4.2\,K) was calculated from the measured critical current at 4.2\,K and the known width and thickness of each line. The straight lines without bends have $j_C$(4.2\,K) = 13.3\,MA/cm$^2$. For the samples with bends the critical current density gradually reduces as $\theta$ increases. For samples with 90$^{\circ}$ bends, the critical current density is 8.1\,MA/cm$^2$, which is approximately 60\% of the value for the straight line. 

Figure ~\ref{fig2} shows the critical current densities $j_C$(4.2\,K) for the samples from batch \#2 with rounded 90-degree bends and different radii of connecting arcs. Although all samples have 90$^{\circ}$ bends, the critical current density increases with the increase in the inner radius $r$ of the connecting arc. At $r$ = 500\,nm, the critical current of the line with bends amounted at almost 80\% of the critical current in the straight line.

\begin{figure}
\includegraphics{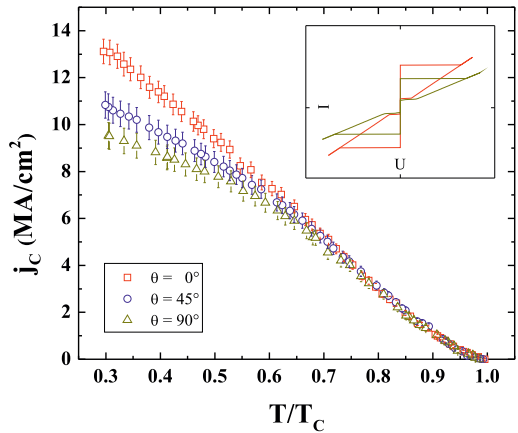}
\caption{\label{fig3}(Color online) Temperature dependences of the critical current densities for three samples with different bending angles. Zero angle corresponds to the straight line. The error of the critical current density is mainly defined by the accuracy of measurement of the width. The inset shows two typical current-voltage curves measured at 4.2\,K for a straight sample (red) and a 90$^{\circ}$ sample (brown).}
\end{figure}

We found that the bends affect the density of the critical current differently at different temperatures. Fig.~\ref{fig3} shows temperature dependences of the critical current density for three samples from the batch \#1 with bending angles 0$^{\circ}$, 45$^{\circ}$ and 90$^{\circ}$. These samples were selected for their smallest difference in the measured width to avoid additional impact of the sample width on $j_C(T)$\cite{1Ilin10,1Ilin12}. The principal behaviour is, however, similar for all samples. In contrast to the significant difference in $j_C$(4.2\,K) the $j_C$ values of all three samples practically coincide in the temperature range from the superconducting transition to approximately 0.7\,$T_C$. At lower temperatures, the $j_C$ values in lines with bends become lower than in the straight lines and the difference grows as the temperature decreases. 

\section{Discussion}

\begin{figure}
\includegraphics{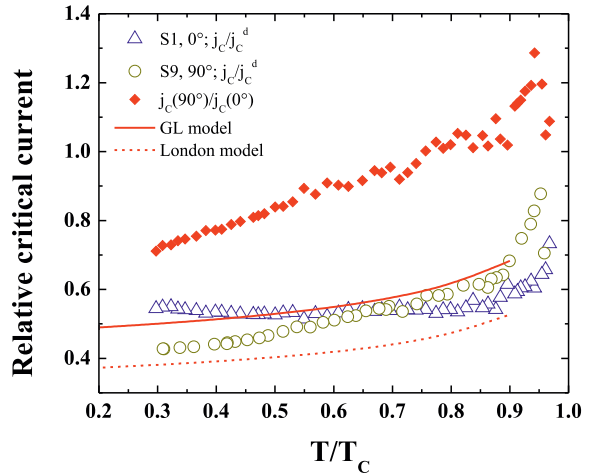}
\caption{\label{fig4}(Color online) Critical current in the line with two sharp 90$^{\circ}$ bends relative to the critical current in the straight line (filled diamonds) and the theoretical predictions of the GL (solid line) and London (dashed line) models. Open triangles and circles show the critical currents in the straight line and in the line with bends, respectively, relative to their depairing critical currents.}
\end{figure}

We compare the experimentally achieved critical current densities in the three samples from the batch \#1 with the depairing current density, which has been computed for each sample in the framework of the Ginzburg-Landau model as
\begin{eqnarray}
\nonumber j_C^d(T)=j_C^B(0)\left[1-\left( \frac{T}{T_C} \right)^2 \right]^{\frac{3}{2}} \\
j_C^B(0)= \frac{4\sqrt{\pi}(e^{\gamma})^2}{21 \zeta(3)\sqrt{3}} \frac{(\Delta(0))^2}{e \rho \sqrt{D \hbar k_B T_C}}
\end{eqnarray}
with the temperature dependence proposed by Bardeen \cite{1Bar62}. The current densities were further corrected for the extreme dirty limit according to the exact calculations of Kupriyanov and Lukichev \cite{1Kup80}. We used the normal state resistivity $\rho$ = 1.84\,\textmu$\Omega$m of our lines at 20\,K along with the electron diffusivity $D = 5\cdot 10^{-5}\,\text{m}^2 \text{s}^{-1}$ and the energy gap at zero temperature $\Delta(0) = 2 k_BT_C$; both are typical for 10\,nm thick NbN films\cite{2Sem09,1Rom04}. The results are shown in Fig.~\ref{fig4} where the ratio of the experimentally measured critical current density to the depairing critical current density $j_C / j_C^d(T)$ is shown as function of temperature for the straight line (triangles) and for the line with two 90$^{\circ}$ bends (circles). It has been found in many experiments that even in straight lines the critical current reaches the depairing value only in the narrow interval below the superconducting transition. At lower temperatures in our straight sample experimental critical current amounts to $\approx$55\% of the depairing value and this ratio remains almost temperature independent at $T < 0.9\,T_C$. The line with the bends also carries an almost depairing supercurrent near the transition, but at low temperatures the ratio gradually decreases from more than 60\% at $T \approx 0.9\,T_C$ to almost 40\% at 0.3\,$T_C$.

\begin{figure}
\includegraphics{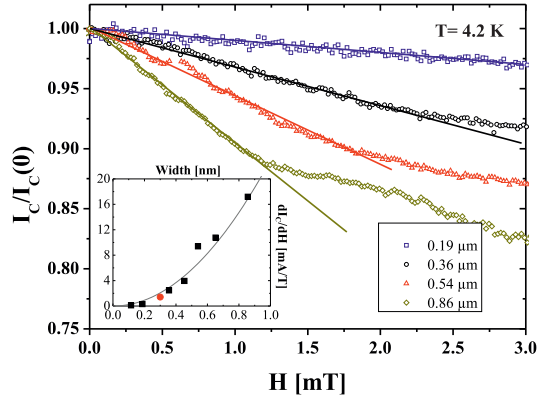}
\caption{\label{fig5}(Color online) Dependence of normalized critical current $I_C/I_C(0)$ on an external magnetic field perpendicular to the sample surface for strips with different width as listed in the legend. The solid lines are linear fits of $I_C(H)$ for $H \rightarrow 0$. The inset shows the dependence of $\text{d}I_C/\text{d}H$ on width of superconducting strips. Symbols are experimental results, the solid line is a $w^2$-fit according to Eq. (2). The red point marks the value for the 300\,nm wide sample from batch \#1 with $\theta=0^{\circ}$.}
\end{figure}

We used the GL equations \cite{1Tin96} to numerically calculate the critical current of the superconducting line with a rectangular cross-section and one sharp 90$^{\circ}$ bend. Simulations were made for the line width $w = 60 \xi_{GL}(0)$ which is close to the experimental situation; in our films the Ginsburg-Landau coherence length at zero temperature $\xi_{GL}(0)$ is approximately 5\,nm \cite{2Sem09}. Since the widths of the studied lines in our experiment were all much smaller then $\lambda_{\text{eff}}$ \cite{1Bar10}, we neglected the screening effects. Although the Likharev's limit is fulfilled almost in the full temperature range except very close to $T_C$, the magnetic field generated by the applied transport current is not strong enough to overcome the edge barrier for vortex penetration \cite{1Ilin10,1Rall11}. We therefore assume that the samples are initially in the Meissner state which becomes unstable at some current which we define as the critical current.
This is confirmed by an analysis of the critical current dependence on external magentic field applied perpendicular to the sample surface. It has been shown theoretically\cite{1Mak98}, that in thin and narrow ($d \ll w \ll \lambda_{\text{eff}}$) superconducting strips at small fields the critical current is
\begin{eqnarray}
I_C(H)=I_C(0)-\frac{w^2}{\lambda_{\text{eff}}}H 
\end{eqnarray}
if the strip is in the Meissner state. Fig. \ref{fig5} shows typical $I_C(H)/I_C(0)$ dependences measured at 4.2\,K for bridges with different widths. The bridges were made from a superconducting film with a thickness similar to the one from which the samples with bends were structured. The solid lines are linear fits of the experimental curves at $H\rightarrow0$. The slope of the curves increases with the width of the bridge. The inset in Fig. \ref{fig5} shows the values of the derivative at the linear part, which is proportional to $w^2$ (solid line) in good agreement with Eq. 2. The red point corresponds to the $\text{d}I_C(H)/\text{d}H$ value obtained on the straight bridge from batch \#1.

The reduction of the critical current in the framework of the London model was computed with Eq. 108 of Ref. 13 describing the sharp bend. The solid and dashed lines in Fig.~\ref{fig4} depict the relative critical current (the ratio of critical currents in the line with bends to that in the straight line) obtained with the Ginzburg-Landau model and the London model (see Eq. 108 of Ref. 13), respectively. The theoretical calculations are limited to temperatures $T < 0.9\,T_C$ since the London model is only valid for line-widths much larger than the coherence length (note that in the GL approach, the theoretical critical current approaches depairing current as $T \rightarrow T_C$). Although both models predict a decrease in the relative critical current at low temperatures, we found that the GL critical current is larger (by a factor of  1.3 for the given width) than it follows from the London model. We believe that the difference appears due to controversy in defining the energy barrier for single vortex entry in the framework of the London model. Indeed, there is a free parameter, which formally describes the minimal distance at which vortex may approach the edge of a straight line. For consistency the minimal distance is defined by setting the barrier to zero for the current equal to the GL depairing current. Although slightly different in different publications, this distance appears comparable with the size of the vortex core. On the other hand, the barrier in the London model can be defined only assuming that this distance is larger than the radius of the core. Our calculations for narrower line ($w = 20 \xi_{GL}(0)$) show that the GL and London critical currents differ by the same factor  $\approx$1.3 only at $T\leq0.6\,T_C$; at larger temperatures this factor decreases and depends on the ratio $w/ \xi_{GL}(T)$.

Numerical and analytical results of the London model \cite{1Cle11}, which we discussed above, demonstrate that a bend becomes active in reducing the critical current when the line width is much larger than the coherence length. Therefore, decreasing the line width in SNSPD meander should improve the spectral range not only due to the decrease in the cross-section of the line but also due to weakening of the effect of bends on the critical current.

Since even in the straight line the measured critical current is smaller than the depairing current, direct comparison of $j_C(T)/j_C^d(T)$ with the theoretical ratio is not wholly appropriate. To estimate the exclusive contribution of the bends, we use the ratio of the experimental critical current in the line with bends to that in the straight line (red diamonds in Fig.~\ref{fig4}). This ratio decreases with the decrease in the temperature and reaches approximately 0.65 at $0.3 T_C$, whereas the theoretical value are both below 0.5 (solid and dashed lines). Assuming that finite rounding may be present even in nominally sharp 90$^{\circ}$ bends and using the expressions for the reduction factor from Ref. 13, we have found negligibly small correction to the critical current for the rounding radius $\approx$25\,nm. This is the largest value of the bend radius which we have found when inspecting lines with sharp bends by the scanning electron microscope. Disagreement with the theory appears most likely due to the difference in the local homogeneity of two samples (the origin of such non-homogeneities is not clear at the moment), which compete with the bends in reducing the critical current. At this point we want to stress that $j_C(90^{\circ})/j_C(0^{\circ})$ values above one at temperatures $T > 0.7\,T_C$ stem from these competing effects between the local inhomogeneities and the rounding of the wires. The bends seem to dominate at $T < 0.7\,T_C$ where the relative critical current approaches theoretical values.

To clarify the matter, further studies should be performed on more uniform (less granular) films from another material. For this case, it has been shown that for straight lines, the depairing critical current can be measured \cite{1Rus04}. Further measurements of the effect of bends on such nanowires may allow for more consistent comparison between the theory and the experiment.

\section{Conclusion}

In summary, we have shown that bends in superconducting lines reduce the experimentally achievable critical current and that the amount of the reduction imposed by particular bend increases with the increase in the bending angle and the decrease in the rounding radius of the bend. In spite of weaker reduction in the critical current predicted by the Ginsburg-Landau model as compared to the London model, we have experimentally found values which are less than expected from any considered model and have different temperature dependence. As the reason for this discrepancy we have proposed local non-homogeneities in the superconducting lines. 

After submission of our paper, a very similar study of critical current in NbTiN films with 90$^{\circ}$ bends was published \cite{1Hor12*}. The found reduction of the critical current was much smaller than the theoretical prediction of the London model \cite{1Cle11} and closer to the result given by the Ginzburg-Landau approach (if one assumes that the same correction factor $\approx$1.3 is valid for wider films than ours).

\begin{acknowledgments}
The work is supported in part by DFG Center for Functional Nanostructures under sub-project A4.3.
\end{acknowledgments}

\end{document}